\documentclass[prd,reprint,showpacs,showkeys,superscriptaddress]{revtex4}
\usepackage{amsfonts}
\usepackage{amssymb}
\usepackage{amsmath}
\usepackage{graphicx}
\usepackage[font={footnotesize,it}]{caption}
\usepackage{latexsym}

\begin{document}

\title{Effect of power-law Maxwell field to the gravitational lensing}
\author{O. Gurtug}
\email{ozaygurtug@maltepe.edu.tr;ozay.gurtug@emu.edu.tr}
\affiliation{T. C. Maltepe University, Faculty of Engineering and Natural Sciences,
Istanbul -Turkey}
\affiliation{Department of Physics, Faculty of Arts and Sciences, Eastern Mediterranean
University, Famagusta, North Cyprus via Mersin 10, Turkey}
\author{M. Mangut}
\email{mert.mangut@emu.edu.tr}
\affiliation{Department of Physics, Faculty of Arts and Sciences, Eastern Mediterranean
University, Famagusta, North Cyprus via Mersin 10, Turkey}

\begin{abstract}
In this paper, we extend the gravitational bending of light studies in
Kottler metrics to comprise nonlinear electrodynamics within the framework
of Einstein - power - Maxwell theory. We show that the closest approach
distance and the gravitational bending of light are affected from the
presence of charge for particular values of the power parameter $k$, which
is defined by means of energy conditions. It is shown that the bending angle
of light is stronger in the case of a strong electric field, which is the
case for $k=1.2$.
\end{abstract}

\maketitle

\section{Introduction}

The question of whether the cosmological constant $\Lambda $ contributes to
the bending angle of light has been pondered by many scientists. The
pioneering study in this regard belongs to N. J. Islam, \ who stated that $%
\Lambda $ has no influence on the bending angle of light \cite{1}. This
result was confirmed by other authors \cite{2,3,4,5,6}. The arguments in 
\cite{1,2,3,4,5,6} are based on the vanishing of the cosmological constant
in the second order null geodesic equation. However, Rindler and Ishak (RI)
have shown that the cosmological constant $\Lambda ,$ does indeed contribute
to the bending angle of light \cite{7}. All these discussions in the
aforementioned papers are based on the Kottler metric \cite{8}, which
describes the geometry of Schwarzschild metric coupled with the cosmological
constant $\Lambda $ (Schwarzschild - de Sitter, SdS)$.$ The marked
distinction between RI and the other authors is the method of calculation of
the bending angle. RI method incorporates with the inner product of two
coordinates in curved space, which paves the way to include the contribution
of all the matter fields existed in the spacetime structure. Therefore, if
one wants to study the effect of the background matter fields on the bending
angle of light, then the method proposed by RI is adequate. Thus, one may
extend the method of RI to include the electric charge together with the
cosmological constant and investigate their combined effect on the
gravitational bending of light.

It has been known from observational stellar data that the compact objects,
namely, Vela X-1, SAXJ1808.4-3658 and 4U1820.30 are categorized as charged
compact stars \cite{9}. The peculiar feature of these compact stars is to
hold a very huge electric charge. The charge value at the surface of the
star is estimated to be $\sim 10^{20}$ Coulomb \cite{10}. Such a huge charge
produces very strong electric field in the surrounding geometry. Solutions
to the Einstein's field equations for a static spherically symmetric systems
have shown that charge associated with massive objects appear as higher
order corrections to the SdS solution. The geometry around the compact
object of such solutions can be associated with the external geometry of a
charged black hole, which may exhibit a region of spacetime filled with
strong electric field in the presence of cosmological constant. From an
astrophysics point of view, it is important to investigate any gravitational
lensing effect that arise due to the presence of charge in addition to the
cosmological constant.

It has been known that the magnetars, which are known as the charged
rotating stars or black holes may produce strong magnetic field. When the
magnetic field is so strong, the standard linear electrodynamics is not a
correct model to describe the geometry around the magnetars. In recent
years, there is a growing interest to use nonlinear electrodynamics in
astrophysics. It has been demonstrated in \cite{11,12} that, unlike the
standard linear Maxwell theory in which the background magnetic field is not
effective on the gravitational redshift, when the background is filled with
nonlinear magnetic field, it contributes to the gravitational redshift. This
contribution is in the sense that, it tends the gravitational redshift to
infinity as the nonlinear magnetic field grows. In analogy to this, if there
is a strong electric field emanated from charged compact stars, its effect
could be studied best by employing nonlinear electrodynamics.

Basically, nonlinear electrodynamics has been introduced to overcome the
divergences in self - energy of point like charges in the standard linear
Maxwell theory. The Born - Infeld nonlinear electrodynamic model was
developed with the expectation to resolve these divergences \cite%
{13,14,15,16,17,18}. It has been shown that this model helps to remove
curvature singularities at the core of black holes \cite{19}.

Another alternative model to nonlinear electrodynamics is the power - law
Maxwell field. In this model, the Lagrangian density of the electromagnetic
field is described by $\mathcal{F}=(F_{\mu \nu }F^{\mu \nu })^{k}$, where $k$
is the nonlinearity parameter. This parameter is a real rational number,
which becomes bounded to some intervals by means of energy conditions. It is
worth to note that, in this model of nonlinear electrodynamics, conformal
invariance condition is satisfied whenever the nonlinearity parameter $k=%
\frac{d}{4}$ is chosen. Here, $d$ denotes the dimension of the spacetime.
This choice implies traceless Maxwell's energy - momentum tensor. In the
last decade, power - law Maxwell field has been used in various studies
ranging from lower to higher dimensions \cite{20,21,22,23,24,25,26,27}.

In the present paper, we shall investigate the effect of nonlinear
electrodynamics on the gravitational bending of light in the presence of
cosmological constant. Because of the observational nature, gravitational
bending of light is the most striking consequence of the Einstein's theory
of relativity. In these phenomena, light emerging from distant
galaxies/stars, bends when it passes near a massive object. There are
considerable amount of research articles that considers the effect of
cosmological constant on the bending angle of light (in addition to
references 1-7, see \cite{31,32}). However, there is no common consensus on
its effect. In this article, we shall go one step forward and investigate
the bending angle of light, when it passes close to a charged compact star
surrounded by strong electric field in the presence of cosmological
constant. This problem is important, because, the existence of neutron stars
or black holes dominated by a strong electric field is a known fact about
our universe. Among the others; Vela X-1, SAXJ1808.4-3658 and 4U1820.30 are
the well known observed charged compact stars (CCS) in astrophysics. In
order to describe the geometry around these CCS in the presence of strong
electric field coupled with the cosmological constant, one may consult
Einstein - power - Maxwell theory that incorporates a nonlinear
electrodynamics through a nonlinear parameter $k$. Within this context, the
solution obtained by Hendi and his co - workers \cite{28} is used for
studying the bending angle of light in the presence of nonlinear
electromagnetic field coupled with the cosmological constant. Though the
contribution of cosmological constant to the bending angle of light has been
extensively studied, the contribution of nonlinear electrodynamics has not
been studied in detail.

The paper is organized as follows. In Sec II, the action of the
Einstein-power-Maxwell formalism and the solution to $\left( 3+1\right) $
dimensional gravity in the presence of cosmological constant is given. The
possible values of nonlinear parameter $k$ is obtained with the help of
energy conditions. The method of calculating the bending angle of light
proposed by RI is briefly explained. In Sec. III, the bending angle of light
is calculated for $k=1$ (which is the linear Maxwell extension of \cite{7}), 
$k=3/4$ and $k=1.2$ (nonlinear Maxwell extension of \cite{7}). The obtained
results are compared with the outcomes of \cite{7} and the contribution of
charge on the bending angle of light is clarified. In section IV, relevant
astrophysical applications are studied numerically for three realistic
charged compact star. The paper is concluded with a results and discussion
in Sec. V.

\section{Einstein - power law Maxwell Field Solutions in $(3+1)-$
Dimensional Gravity}

The $\left( 3+1\right) -$dimensional action in Einstein - power law Maxwell
theory of gravity with a cosmological constant $\Lambda $ is given by,%
\begin{equation}
I=-\frac{1}{16\pi }\int d^{4}x\sqrt{-g}\left\{ R-2\Lambda +\mathcal{L}\left( 
\mathcal{F}\right) \right\} ,
\end{equation}%
in which $R$ is the Ricci scalar, $\Lambda =\frac{3}{l^{2}}$ is the positive
cosmological constant (for asymptotically de-Sitter solutions) with a length
scale $l$ and $\mathcal{L}\left( \mathcal{F}\right) =-\left\vert \mathcal{F}%
\right\vert ^{k}$ where $k$ is the nonlinearity parameter with the Maxwell
invariant $\mathcal{F=}F_{\mu \nu }F^{\mu \nu }$. Note that linear Maxwell
limit is restored when $k=1$. The metric ansatz for $\left( 3+1\right) -$
dimensional gravity is given in standard form by%
\begin{equation}
ds^{2}=-f(r)dt^{2}+\frac{dr^{2}}{f(r)}+r^{2}\left( d\theta ^{2}+\sin
^{2}\theta d\varphi ^{2}\right) .
\end{equation}%
The solution to the Einstein-power law Maxwell equations was given in any
dimension in \cite{28}, and the particular solution in $\left( 3+1\right) -$
dimensional gravity is given by 
\begin{equation}
f(r)=1-\frac{\Lambda r^{2}}{3}-\frac{m}{r}+\left\{ 
\begin{array}{cc}
\frac{2^{3/2}q^{3}}{r}\ln \left( \frac{r}{l}\right) ,\text{ } & k=\frac{3}{2}%
, \\ 
\frac{\left( 2k-1\right) ^{2}\left( \frac{2\left( 2k-3\right) ^{2}q^{2}}{%
\left( 2k-1\right) ^{2}}\right) ^{k}}{2\left( 3-2k\right) r^{2/\left(
2k-1\right) }}, & otherwise,\ \ except\text{ \ }for\text{ }k\neq \frac{1}{2},%
\end{array}%
\right.
\end{equation}%
in which $q$ and $m$ are charge and mass related integration constants. The
electric charge $Q$ and the ADM mass $M$ of the object are defined by,%
\begin{equation}
M=\frac{m}{2},
\end{equation}

\begin{equation}
Q=\left\{ 
\begin{array}{cc}
\frac{3}{4\sqrt{2}}q^{2}, & k=\frac{3}{2}, \\ 
\frac{k\left( 2k-1\right) }{\sqrt{2}}2^{k-1/2}\left( \frac{\left(
3-2k\right) q}{2k-1}\right) ^{2k-1},\text{ } & otherwise,\text{ \ }except%
\text{ }for\text{\ }k\neq \frac{1}{2},%
\end{array}%
\right. .
\end{equation}

\subsection{Energy Conditions}

Before calculating the bending angle of light in the presence of nonlinear
electrodynamics, the energy conditions must be checked for possible values
of the parameter $k$. This is important within the context of the considered
model of nonlinear electrodynamics, as far as the physically acceptable
solutions are concerned.

The energy momentum tensor of the power - law Maxwell field is given by, 
\begin{equation}
T_{\mu }^{\nu }=\frac{1}{2}\left\{ \mathcal{L(F)}\text{ }\mathcal{\delta }%
_{\mu }^{\nu }-4\mathcal{L}_{\mathcal{F}}\mathcal{(F)}\left( F_{\mu \lambda
}F^{\nu \lambda }\right) \right\} ,
\end{equation}%
in which $\mathcal{L}_{\mathcal{F}}\mathcal{(F)=}\frac{\partial \mathcal{L(F)%
}}{\partial \mathcal{F}}.$ The nonzero component of the electromagnetic
field tensor $F_{\mu \nu }=F_{tr}$ is given by%
\begin{equation}
F_{tr}=\left\{ 
\begin{array}{cc}
-\frac{q}{r}\text{ ,\ \ } & k=\frac{3}{2}, \\ 
\frac{k\left( 2k-1\right) }{\sqrt{2}}2^{k-1/2}\left( \frac{\left(
3-2k\right) q}{2k-1}\right) ^{2k-1}r^{-\left( \frac{2}{2k-1}\right) },\text{ 
} & otherwise,\text{ \ \ }except\text{ }for\text{ }k\neq \frac{1}{2},%
\end{array}%
\right. .
\end{equation}%
As a direct consequence, the Maxwell invariant $\mathcal{F=}F_{\mu \nu
}F^{\mu \nu }=-2\left( F_{tr}\right) ^{2}=-2\left( E\right) ^{2}$, where $E$
is the electric field.

The weak energy conditions (WEC) state that%
\begin{equation}
\rho \geq 0,\text{\ \ \ \ \ \ }\rho +p_{r}\geq 0,\text{\ \ \ }\rho
+p_{\theta }\geq 0\text{\ \ and \ \ \ }\rho +p_{\varphi }\geq 0,
\end{equation}%
where $\rho $ is the energy density, $p_{r},$ $p_{\theta }$ and $p_{\varphi
} $ are the principal pressures defined by,%
\begin{equation}
\rho =-T_{t}^{t}=-\frac{1}{2}\left( 2k-1\right) \mathcal{F}^{k},
\end{equation}%
\begin{equation}
p_{r}=T_{r}^{r}=\frac{1}{2}\left( 2k-1\right) \mathcal{F}^{k},
\end{equation}%
\begin{equation}
p_{\theta }=T_{\theta }^{\theta }=T_{\varphi }^{\varphi }=p_{\varphi }=-%
\frac{1}{2}\mathcal{F}^{k}.
\end{equation}%
WEC is satisfied whenever $k>\frac{1}{2}.$ The strong energy condition (SEC)
states that%
\begin{equation}
\rho +\sum_{i=1}^{3}p_{i}\geq 0\text{ , \ \ \ \ \ }\rho +p_{r}\geq 0,\text{\
\ \ }\rho +p_{\theta }\geq 0\text{\ \ and \ \ \ }\rho +p_{\varphi }\geq 0%
\text{\ .\ \ }
\end{equation}%
This condition together with the WEC reveals that $k>\frac{1}{2}.$ The
dominant energy condition (DEC) states that 
\begin{equation}
p_{eff}=\frac{1}{2}\sum_{i=1}^{3}T_{i}^{i}\geq 0.
\end{equation}%
This condition yields $k\leq \frac{3}{2}.$ If WEC, SEC and DEC are combined, 
$k,$ gets bounded to $\frac{1}{2}<k\leq \frac{3}{2}.$ In addition to energy
conditions, one can also impose the causality condition which is defined by%
\begin{equation}
0\leq \frac{p_{eff}}{\rho }\leq 1.
\end{equation}%
The analysis has revealed that the causality condition is satisfied for $%
\frac{1}{2}<k\leq \frac{3}{2}.$ As a consequence, if the nonlinearity
parameter $k$ is chosen such that it satisfies the constraint condition $%
\frac{1}{2}<k\leq \frac{3}{2},$ then all the energy conditions are satisfied
and the resulting solution to the Einstein-power law Maxwell equations
becomes physically acceptable.

\subsection{Bending Angle}

As is well known, the inner product of two vectors remains invariant under
the rotation of coordinate systems. Rindler and Ishak have used this
property in \cite{7} to calculate the relativistic bending angle of light in
the following way. The angle between two coordinate directions $d$ and $%
\delta $ as shown in Fig.1 is given by the invariant formula,

\begin{equation}
\cos \left( \psi \right) =\frac{d^{i}\delta _{i}}{\sqrt{\left(
d^{i}d_{i}\right) \left( \delta ^{j}\delta _{j}\right) }}=\frac{%
g_{ij}d^{i}\delta ^{j}}{\sqrt{\left( g_{ij}d^{i}d^{j}\right) \left(
g_{kl}\delta ^{k}\delta ^{l}\right) }}.
\end{equation}%
In this formula, $g_{ij}$ is the metric tensor of the constant time slice of
the metric (2), a two-dimensional curved $(r,\varphi )$ space, which is
defined at the equatorial plane (when $\theta =\pi /2$ ) in the following
form \cite{29}, as the orbital plane of the light rays , 
\begin{equation}
dl^{2}=\frac{dr^{2}}{f(r)}+r^{2}d\varphi ^{2}.
\end{equation}%
As a requirement of the formalism, we need to define null geodesics
equation. The constants of motion in the considered spacetime are%
\begin{equation}
\frac{dt}{d\tau }=-\frac{E}{f(r)},\text{ \ \ \ \ \ }\frac{d\varphi }{d\tau }=%
\frac{h}{r^{2}},\text{ \ }
\end{equation}%
in which $\tau $ stands for proper time. Using these conserved quantities,
we obtain%
\begin{equation}
\left( \frac{dr}{d\tau }\right) ^{2}=E^{2}-\frac{h}{r^{2}}f(r),
\end{equation}%
and%
\begin{equation}
\left( \frac{dr}{d\varphi }\right) ^{2}=\frac{r^{4}}{h^{2}}\left( E^{2}-%
\frac{h^{2}}{r^{2}}f(r)\right) ,
\end{equation}%
where $E$ and $h$ represent energy and angular momentum, respectively. It
has been found convenient to introduce a new variable $u$, such that, $u=%
\frac{1}{r}.$ \ Using this transformation, Eq.(19) transforms to%
\begin{equation}
\frac{d^{2}u}{d\varphi ^{2}}=-uf(u)-\frac{u^{2}}{2}\frac{df(u)}{du}.
\end{equation}%
Once the above differential equation is solved, the obtained solution is
used to define another equation in the following way,%
\begin{equation}
A(r,\varphi )\equiv \frac{dr}{d\varphi }.
\end{equation}%
Now, if the direction of the orbit is denoted by $d$ and that of the
coordinate line $\varphi =$ constant $\delta ,$ we have%
\begin{eqnarray}
d &=&\left( dr,d\varphi \right) =\left( A,1\right) d\varphi \text{ \ \ \ \ \
\ \ }d\varphi <0,  \notag \\
\delta &=&\left( \delta r,0\right) =\left( 1,0\right) \delta r.
\end{eqnarray}%
If we use these definitions in (15), we obtain,

\begin{equation}
\tan \left( \psi \right) =\frac{\left[ g^{rr}\right] ^{1/2}r}{\left\vert
A(r,\varphi )\right\vert }.
\end{equation}%
The one - sided bending angle is therefore defined as $\epsilon =\psi
-\varphi .$

\section{Bending of Light in the Presence of Linear and Nonlinear
Electrodynamics}

The main purpose of this paper is to study the effect of linear and
nonlinear electromagnetic fields ( in the form of power-law Maxwell
invariant described by $(F_{\mu \nu }F^{\mu \nu })^{k}$, where $k$ is the
nonlinearity parameter.) on the bending angle of light. Our motivation for
introducing nonlinear electrodynamics is as follows: When the light passes
through a region in which the surrounding geometry is filled by strong
electric field, such a strong electric field is best described by nonlinear
electrodynamics. This effect will be investigated in $\left( 3+1\right) -$
dimensional geometry where the power - law Maxwell field is coupled to
Schwarzschild- de Sitter (SdS) metric. In the present paper, we shall
consider the extension of the RI's paper with different values of parameter $%
k$. We shall investigate the cases where $k=1$ (linear electrodynamics), $%
k=3/4$ ($k<1$) and $k=1.2$ ($k>1$) ( nonlinear electrodynamics ). 
\begin{figure}[h]
\includegraphics[width=120mm,scale=1]{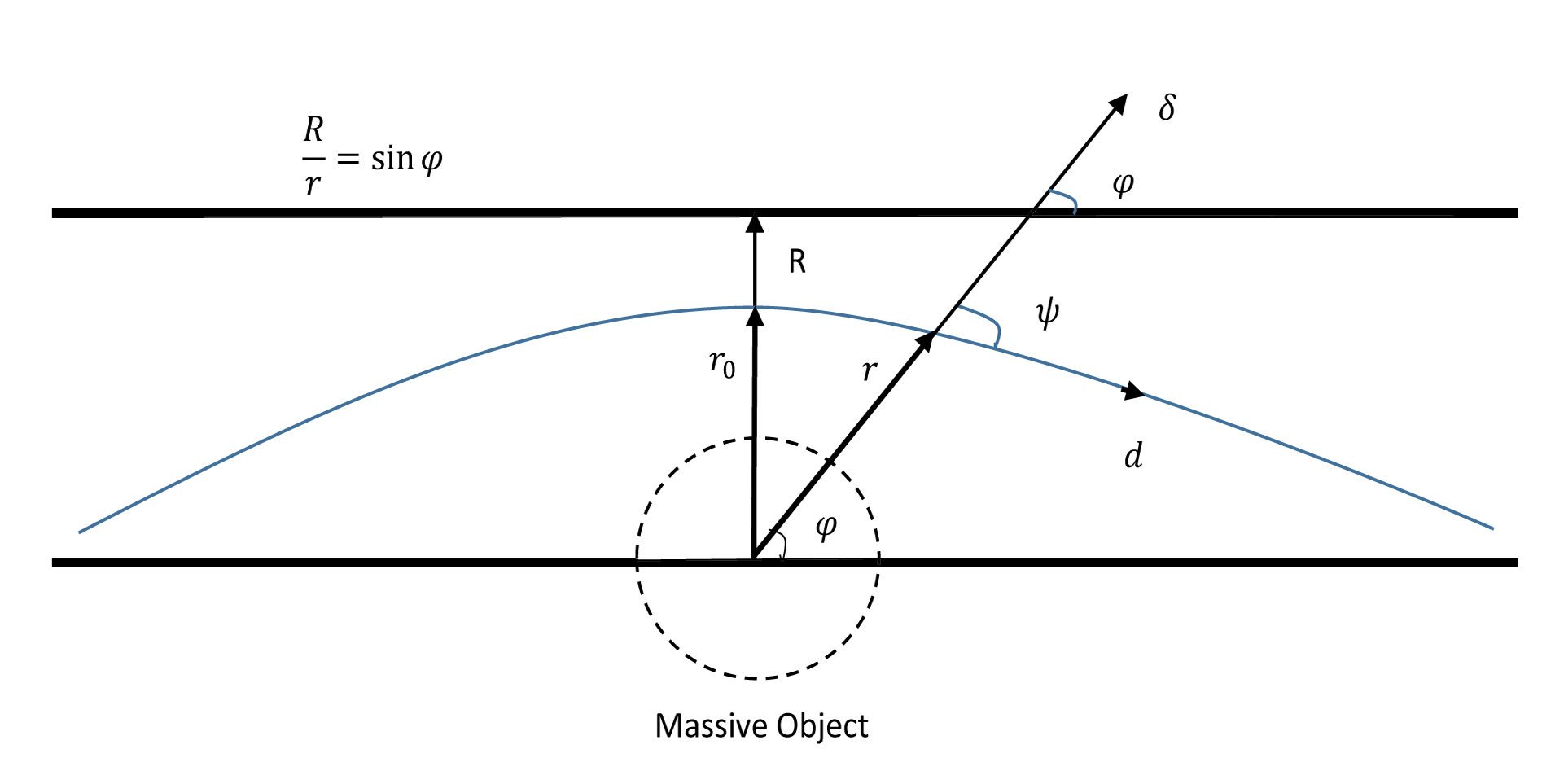}
\caption{A diagram of light bending in the presence of a massive object.}
\end{figure}

\subsection{ The case in linear electrodynamics: $k=1$}

In this subsection, we will extend the study of RI for the SdS case to the
charged SdS. This problem has already been considered in \cite{30} partly,
by employing the method of RI. The contribution of electric charge to the
bending angle of light within the context of Reissner-Nordstr\"{o}m - de
Sitter metric is shown. In the present paper, the effect of the electric
charge and the cosmological constant on the bending of light will be
investigated in more detail. The spacetime geometry for this case is
described by%
\begin{equation}
f(r)=1-\frac{m}{r}-\frac{\Lambda r^{2}}{3}+\frac{Q^{2}}{r^{2}}.
\end{equation}%
Here $Q=q$. The orbital equation for the light in this spacetime is obtained
from Eq.(20), and is given by%
\begin{equation}
\frac{d^{2}u}{d\varphi ^{2}}+u=\frac{3}{2}mu^{2}-2Q^{2}u^{3}.
\end{equation}%
The homogeneous part of equation (25) has solution in harmonic form. At this
stage, we prefer to use the same solution used in \cite{7}, namely \ $\frac{%
\sin \varphi }{R}.$ This solution corresponds to the undeflected light in
the absence of gravity, displayed as a solid horizontal line in Fig. 1. This
choice ensures that we recover the results found in \cite{7}, when we set $%
Q=0$. \ Then, we substitute the first order homogeneous solution to the
right hand side and solve for the full inhomogeneous equation (25), which
admits the approximate solution as%
\begin{equation}
u=\frac{1}{r}=\frac{\sin \varphi }{R}+\frac{1}{4R^{3}}\left\{ 2mR\left(
1+\cos ^{2}\varphi \right) +Q^{2}\left( 3\varphi \cos \varphi -\sin \varphi
\cos ^{2}\varphi -2\sin \varphi \right) \right\} .
\end{equation}%
We differentiate Eq.(26) with respect to $\varphi ,$ in accordance with
Eq.(21) to get $\ A(r,\varphi ),$ 
\begin{equation}
A(r,\varphi )=\frac{r^{2}}{4R^{3}}\left\{ 2mR\sin 2\varphi +Q^{2}\left( \cos
^{3}\varphi -\sin \varphi \sin 2\varphi +3\varphi \sin \varphi -\cos \varphi
\right) \right\} -\frac{r^{2}}{R}\cos \varphi .
\end{equation}%
In equations (26) and (27), the constant parameter $R$ is called the impact
parameter and in the case of aymptotically flat metrics it is defined as $b$%
. As mentioned in \cite{7}, since the considered spacetime is not
asymptotically flat, the effect of other parameters should also be taken
into account. Hence, in conjunction with \cite{7}, this parameter is related
with the physically meaningful area distance $r_{0\text{ }}$of closest
approach by,%
\begin{equation}
\frac{1}{r_{0}}=\frac{1}{R}+\frac{m}{2R^{2}}-\frac{Q^{2}}{2R^{3}}.
\end{equation}%
From this result, it is seen that the closest approach distance increases
when compared to the uncharged case \cite{7}. Note that, the cosmological
constant $\Lambda $ does not have any contribution to the closest distance $%
r_{0}$.

The one - sided bending angle $\epsilon $ of light is calculated by using
Eq.(23). As can be seen from Fig.1, the value of this angle is measured
relative to the coordinate planes where $\varphi =$ constant. For the small
bending angle, $\tan \psi _{0}\approx \psi _{0}.$ We then take $\varphi =0,$
for large distance away from the source. For this particular case, the one -
sided bending angle is%
\begin{equation}
\epsilon =\psi _{0}=\frac{m}{R}\left\{ 1-\frac{\Lambda R^{4}}{3m^{2}}-\frac{%
m^{2}}{R^{2}}+\frac{Q^{2}m^{2}}{R^{4}}\right\} ^{1/2}\simeq \frac{m}{R}%
\left\{ 1-\frac{\Lambda R^{4}}{6m^{2}}-\frac{m^{2}}{2R^{2}}+\frac{Q^{2}m^{2}%
}{2R^{4}}\right\} +\mathcal{O}\left( \frac{Q^{4}m^{5}}{R^{9}}\right) .
\end{equation}%
The total bending angle is defined as the twice of this angle, namely, $%
2\psi _{0}.$ It is important to note the difference in the contribution to
the bending angle of light between the cosmological constant and the
electric charge. While the positive cosmological constant decreases the
bending angle, the electric charge has the tendency to increase it. As an
observational viewpoint this contribution may be negligibly small, but from
the theoretical viewpoint it is important to see how the electric charge
enters the calculation.

In order to explore the contribution of electric charge in the presence of
the cosmological constant, we consider also the bending angle occurring at $%
\varphi =\pi /4$, rather than zero. This value is chosen intentionally to
compare the obtained results with the outcomes of RI's work \cite{7}. When $%
\varphi =\pi /4$ in Eq.(26), we have,%
\begin{equation}
r=\frac{4R^{3}}{2\sqrt{2}R^{2}+3mR+\frac{Q^{2}}{2\sqrt{2}}\left( \frac{3\pi 
}{2}-5\right) }.
\end{equation}%
If we assume that $\frac{m}{R}$ $\ll 1$ and $\Lambda R^{2}\ll 1$ as in \cite%
{7}, we obtain,%
\begin{equation}
r=\sqrt{2}R,\text{ \ \ \ \ \ \ \ }A(r,\pi /4)=-\sqrt{2}R\left( 1-\frac{m}{%
\sqrt{2}R}\right) ,
\end{equation}%
\begin{equation}
\tan \left( \psi \right) =1+\frac{m}{2\sqrt{2}R}-\frac{\Lambda R^{2}}{3}+%
\frac{Q^{2}}{4R^{2}}.
\end{equation}%
Note that the one - sided bending angle is defined as $\epsilon =\psi
-\varphi $ and for small angle it may be written as, $\epsilon \simeq \tan
\left( \psi -\varphi \right) =\frac{\tan \psi -\tan \varphi }{1+\tan \psi
\tan \varphi }.$ Since $\tan \varphi =1,$ we obtain the one - sided bending
angle as,%
\begin{equation}
\epsilon =\frac{m}{4\sqrt{2}R}-\frac{\Lambda R^{2}}{6}+\frac{Q^{2}}{8R^{2}}.
\end{equation}%
This result indicates that the effect of cosmological constant $\left( \text{%
when, }\Lambda >0\right) $ and the electric charge on the bending angle of
light is not in phase. Furthermore, the contribution of the charge to the
bending angle of light is more dominant when compared to small angle
calculation ($\psi _{0}$, namely Eq.(29)). Of course, the above result is a
consequence of the assumption made on the values of $\frac{m}{R}$ $\ll 1$
and $\Lambda R^{2}\ll 1.$ The exact results without imposing these
conditions are as follows:%
\begin{equation}
A(r,\pi /4)=\frac{r^{2}}{4R^{3}}\left\{ 2mR+\frac{3Q^{2}}{4\sqrt{2}}\left(
\pi -2\right) -2\sqrt{2}R^{2}\right\}
\end{equation}%
and%
\begin{equation}
\tan \left( \psi \right) =\frac{4R^{3}\left( 1-\frac{m}{r}-\frac{\Lambda
r^{2}}{3}+\frac{Q^{2}}{r^{2}}\right) ^{1/2}}{r\left\vert 2mR+\frac{3Q^{2}}{4%
\sqrt{2}}\left( \pi -2\right) -2\sqrt{2}R^{2}\right\vert }
\end{equation}%
where $r$ is given in Eq.(30), and the one-sided bending angle becomes,%
\begin{equation}
\epsilon \simeq \tan \left( \psi -\varphi \right) =\frac{\tan \left( \psi
\right) -1}{1+\tan \left( \psi \right) }.
\end{equation}

\subsection{The case in nonlinear electrodynamics: $k=\frac{3}{4}$}

The metric in this case is given by%
\begin{equation}
f(r)=1-\frac{m}{r}-\frac{\Lambda r^{2}}{3}+\frac{\tilde{Q}}{r^{4}}
\end{equation}%
in which $\ \tilde{Q}$ is related to the star's charge $Q$ through, $\tilde{Q%
}=\frac{\left( 18q^{2}\right) ^{3/4}}{12}=4.469Q^{3}.$ \ The orbital
equation of the light is obtained as%
\begin{equation}
\frac{d^{2}u}{d\theta ^{2}}+u=\frac{3}{2}mu^{2}-3\tilde{Q}u^{5}.
\end{equation}%
The approximate solution of this equation is found to be%
\begin{equation}
u=\frac{1}{r}=\frac{\sin \varphi }{R}+\frac{1}{48R^{5}}\left\{ \tilde{Q}%
\left( \sin 2\varphi \cos ^{3}\varphi -\frac{9}{2}\sin 2\varphi \cos \varphi
-8\sin \varphi \right) +24mR^{3}\left( 1+\cos ^{2}\varphi \right) \right\} ,
\end{equation}%
and equation (21) becomes,%
\begin{multline}
A(r,\varphi )=\frac{\tilde{Q}r^{2}}{48R^{5}}\left( 2\sin ^{2}2\varphi \cos
\varphi +9\cos ^{3}\varphi +15\varphi \sin \varphi -2\cos ^{5}\varphi -7\cos
\varphi -9\sin 2\varphi \sin \varphi \right) + \\
\frac{r^{2}}{R}\left( \frac{m}{2R}\sin 2\varphi -\cos \varphi \right) .
\end{multline}%
The closest distance of approach $r_{0},$ in the presence of nonlinear
electrodynamics becomes, 
\begin{equation}
\frac{1}{r_{0}}=\frac{1}{R}+\frac{m}{2R^{2}}-\frac{\tilde{Q}}{6R^{5}}.
\end{equation}%
When we compare equations (28) and (41), it is observed that the closest
distance decreases with respect to the linear Maxwell case. Next, we
calculate the bending angle when $\varphi =0,$ which is the bending angle
named as the small angle $\psi _{0}.$ For this particular case we found that 
\begin{equation}
r=\frac{R^{2}}{m},\text{ \ \ \ \ }A(r,0)=-\frac{R^{3}}{m^{2}}\text{\ ,}
\end{equation}%
then the one - sided bending angle becomes%
\begin{equation}
\epsilon =\psi _{0}=\frac{m}{R}\left\{ 1-\frac{\Lambda R^{4}}{3m^{2}}-\frac{%
m^{2}}{R^{2}}+\frac{\tilde{Q}m^{4}}{R^{6}}\right\} ^{1/2}\simeq \frac{m}{R}%
\left\{ 1-\frac{\Lambda R^{4}}{6m^{2}}-\frac{m^{2}}{2R^{2}}+\frac{\tilde{Q}%
m^{4}}{2R^{6}}\right\} +\mathcal{O}\left( \frac{\tilde{Q}^{2}m^{5}}{R^{13}}%
\right) .
\end{equation}%
This result indicates that the contribution of the charge to the bending
angle of light is negligible, due to the fact that $\frac{m}{R}\ll 1.$ For
the sake of completeness, it is of interest to look at the bending angle of
light when $\varphi =\pi /4.$ The values of $r$ and $A(r,\pi /4)$ is exactly
the same as in Eq.(31), while $\tan \left( \psi \right) $ is obtained as, 
\begin{equation}
\tan \left( \psi \right) =1+\frac{m}{2\sqrt{2}R}-\frac{\Lambda R^{2}}{3}+%
\frac{\tilde{Q}}{8R^{4}},
\end{equation}%
we find the one - sided bending angle of light as 
\begin{equation}
\epsilon =\frac{m}{4\sqrt{2}R}-\frac{\Lambda R^{2}}{6}+\frac{\Lambda \tilde{Q%
}}{48R^{2}}.
\end{equation}%
Note that, in obtaining the Eq. (45), only the dominant terms are preserved,
the higher order terms are ignored. The peculiar feature of nonlinear
electrodynamics is very clear in the above equation. The charge and the
cosmological constant are coupled together.

\subsection{The case in nonlinear electrodynamics: $k=1.2$}

In this subsection, we consider the case where the nonlinearity parameter $%
k>1.$ The solution for this particular case describes a region of spacetime,
which is dominated by strong electric field. The bending angle of light is
calculated for $k=1.2$. The metric function for the power parameter $k=1.2$
is obtained from Eq.(3) which yields,%
\begin{equation}
f(r)=1-\frac{\Lambda r^{2}}{3}-\frac{m}{r}+\frac{0.484Q^{12/7}}{r^{10/7}}.
\end{equation}%
The equation for the light in this spacetime is obtained from Eq.(20) as,%
\begin{equation}
\frac{d^{2}u}{d\varphi ^{2}}+u=\frac{3m}{2}u^{2}-0.824Q^{12/7}u^{17/7}.
\end{equation}%
The first approximate solution $u=\frac{\sin \varphi }{R},$ is substituted
back in Eq.(47) and its resulting solution for $u$ is obtained as%
\begin{equation}
u=\frac{1}{r}=\frac{\sin \varphi }{R}+\frac{m}{2R^{2}}\left( \cos
^{2}\varphi +1\right) -\frac{0.484Q^{12/7}}{R^{17/7}}\left\{ \frac{7}{24}%
\sin ^{31/7}\varphi -\cos \varphi \int \sin ^{24/7}\varphi d\varphi \right\}
,
\end{equation}%
and the equation (21) becomes,%
\begin{equation}
A(r,\varphi )=-r^{2}\left\{ \frac{\cos \varphi }{R}-\frac{m}{2R^{2}}\sin
2\varphi -\frac{0.484Q^{12/7}}{R^{17/7}}\left[ \frac{31}{24}\cos \varphi
\sin ^{24/7}\varphi +\sin \varphi \int \sin ^{24/7}\varphi d\varphi -\cos
\varphi \sin ^{24/7}\varphi \right] \right\}
\end{equation}%
The integral expression in equations (48) and (49), whenever necessary can
be evaluated in terms of incomplete Beta functions . The closest approach
distance $r_{0}$\ occurs when $\varphi =\pi /2,$ which is found to be%
\begin{equation}
\frac{1}{r_{0}}=\frac{1}{R}+\frac{m}{2R^{2}}-\frac{0.625Q^{12/7}}{R^{17/7}}.
\end{equation}%
\ The comparison of the closest approach distance to the results found
formerly for $k=3/4$ and $k=1$ reveals that when $k=1.2,$ the closest
approach distance $r_{0}$ becomes larger than the other two cases. The one -
sided bending angle measured at $\varphi =0$ is given by%
\begin{multline}
\epsilon =\psi _{0}=\frac{m}{R}\left\{ 1-\frac{\Lambda R^{4}}{3m^{2}}-\frac{%
m^{2}}{R^{2}}+\frac{0.484Q^{12/7}m^{10/7}}{R^{20/7}}\right\} ^{1/2}\simeq \\
\frac{m}{R}\left\{ 1-\frac{\Lambda R^{4}}{6m^{2}}-\frac{m^{2}}{2R^{2}}+\frac{%
0.242Q^{12/7}m^{10/7}}{R^{20/7}}\right\} +\mathcal{O}\left( \frac{%
Q^{24/7}m^{27/7}}{R^{47/7}}\right) .
\end{multline}

The calculation of one-sided bending angle for three different $k$
parameters indicates that the charge of the compact star contributes to the
bending angle. In contrast to the positive cosmological constant, the charge
of the star has the tendency to increases the bending angle of light. The
next section is devoted to discuss numerically about the effect of power
parameter $k$ and the electric charge $Q,$ by using the real approximate
values of three different charged compact stars. 
\begin{figure}[h]
\includegraphics[width=160mm,scale=1]{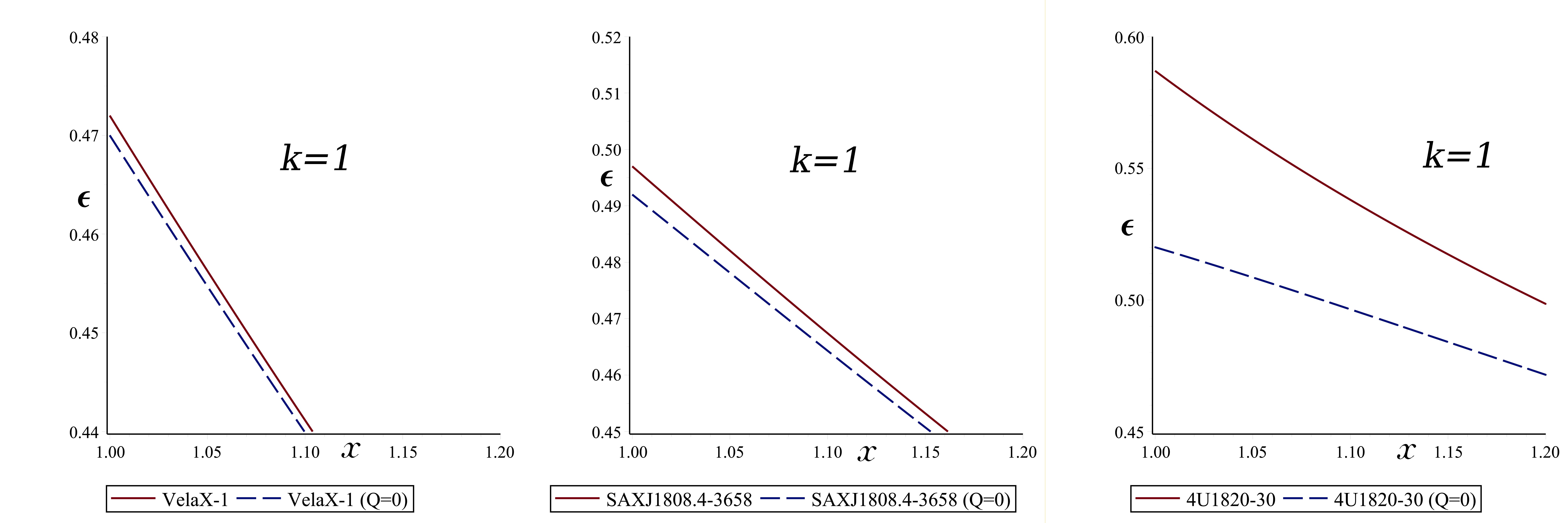}
\caption{The bending angle $\protect\epsilon $ versus $x,$ for the linear
case $k=1$, have been plotted for the compact objects Vela X-1,
SAXJ1808.4-3658 and 4U1820.30. The solid line indicate the variation in the
bending angle as the parameter $x=R/R_{\ast }$, ( here $R_{\ast }$ denotes
the radius of the charged compact star) changes. The dashed line represent
the variation in the absence of charge.Fig2a, 2b and 2c belongs to Vela X-1,
SAXJ1808.4-3658 and 4U1820.30, respectively. It is important to emphasize
that the possible pair creation near the surface of the compact objects are
ignored. The above figures display only the behaviour of the bending angle
as the distance parameter $x$ increases with and without charge.}
\end{figure}
\begin{figure}[h]
\includegraphics[width=160mm,scale=1]{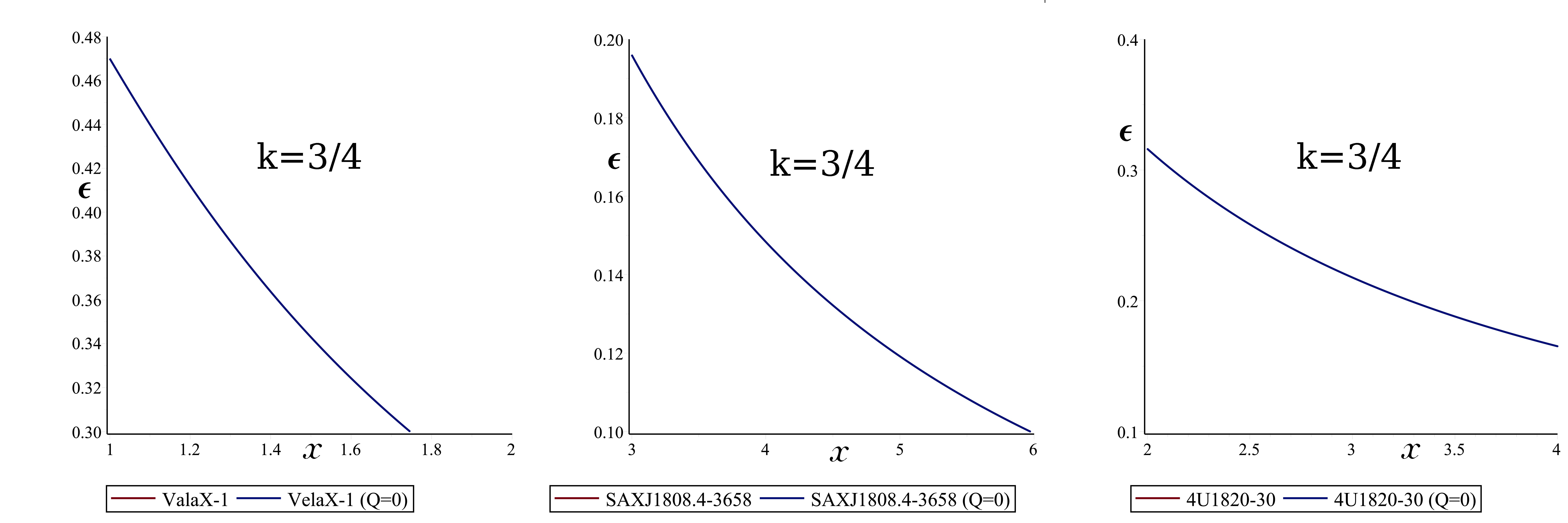}
\caption{The bending angle $\protect\epsilon $ versus $x,$ for the nonlinear
electrodynamic case when $k=3/4<1$, have been plotted for the compact
objects Vela X-1, SAXJ1808.4-3658 and 4U1820.30. The effect of charge is
almost negligible and curves with and without charge coincides with each
other. Fig2a, 2b and 2c belongs to Vela X-1, SAXJ1808.4-3658 and 4U1820.30,
respectively.}
\end{figure}
\begin{figure}[h]
\includegraphics[width=160mm,scale=1]{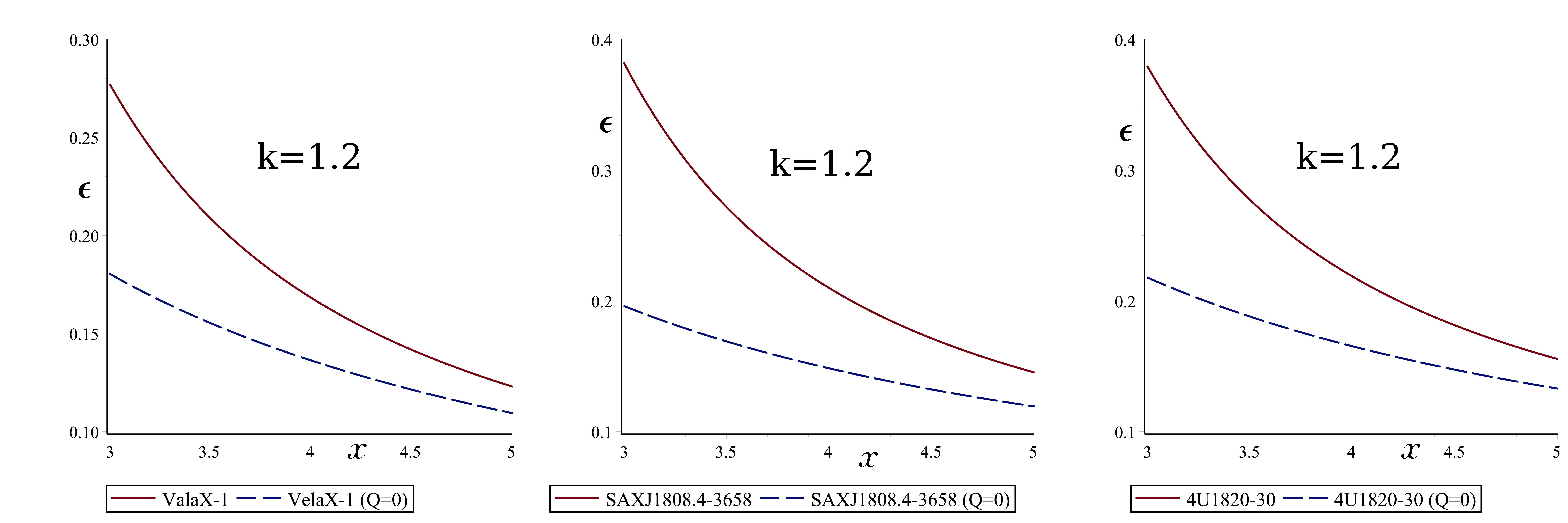}
\caption{The bending angle $\protect\epsilon $ versus $x,$ for the nonlinear
case when $k=1.2>1$, have been plotted for the compact objects Vela X-1,
SAXJ1808.4-3658 and 4U1820.30. The solid line indicate the variation in the
bending angle as the parameter $x=R/R_{\ast }$, ( here $R_{\ast }$ denotes
the radius of the charged compact star) changes. The dashed line represent
the variation in the absence of charge.Fig4a, 4b and 4c belongs to Vela X-1,
SAXJ1808.4-3658 and 4U1820.30, respectively. As in the case of $k=1$, it is
important to emphasize that the possible pair creation near the surface of
the compact objects are ignored. The above figures display only the
behaviour of the bending angle as the distance parameter $x$ increases with
and without charge.}
\end{figure}
\begin{figure}[h]
\includegraphics[width=160mm,scale=1]{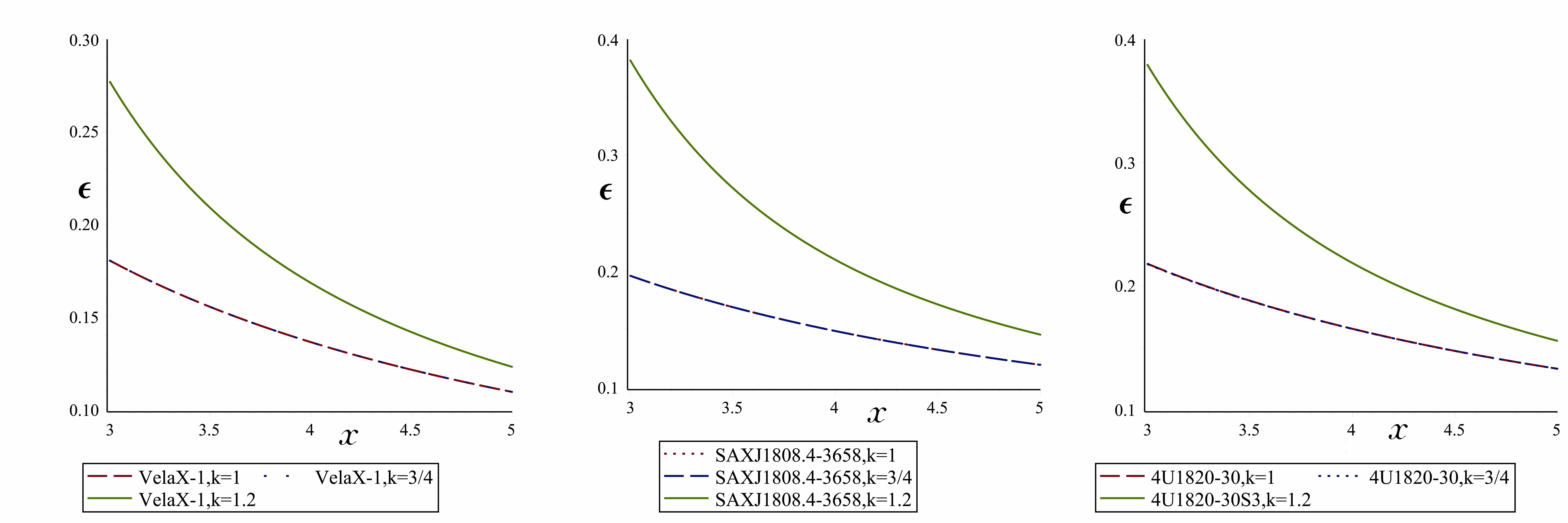}
\caption{The bending angle $\protect\epsilon $ versus $x,$ have been plotted
for the charged compact objects Vela X-1, SAXJ1808.4-3658 and 4U1820.30.
Figures display, how the bending angle of light affected when the power
parameter $k$ changes. Fig5a, 5b and 5c are for Vela X-1, SAXJ1808.4-3658
and 4U1820.30, respectively. It should be noted that for each of the compact
object, the behaviour of the bending angle for $k=1$ and $k=3/4$ is almost
the same, thus the corresponding curves coincides with each other.}
\end{figure}

\section{Relevant Astrophysical Applications}

In this section, we discuss relevant astrophysical applications. The
obtained bending angles for different power parameter $k$ are studied
numerically to display the effect of electric charge in the presence of
cosmological constant. Our numerical analysis are carried for three
realistic charged compact stars whose properties are tabulated in Table-1 
\cite{9}.

\begin{table}[th]
\caption{The approximate values of the masses, radii and charges of the
charged compact stars. Here $M_{\odot }$ denotes the mass of the sun. }
\label{table:nonlin}
\centering
\begin{tabular}{|l|c|c|r|}
\hline
Charged Compact Stars & M & Radius (km) & Electric Charge (C) \\ \hline\hline
Vela X-1 (CS1) & $1.77M_{\odot }$ & $9.56$ & $1.81\times 10^{20}$ \\ \hline
\multicolumn{1}{|c|}{SAXJ 1808.4-3658 (CS2)} & $1.435M_{\odot }$ & $7.07$ & $%
1.87\times 10^{20}$ \\ \hline
4U 1820-30 (CS3) & $2.25M_{\odot }$ & $10$ & $1.89\times 10^{20}$ \\ \hline
\end{tabular}
\end{table}

In our numerical analysis, we take $\varphi =0$ as the reference point at
which the one-sided bending angle is measured. This point corresponds to a
very large distance away from the source. The bending angle $\epsilon $ is
plotted against $x=R/R_{\ast }$, here $R_{\ast }$ denotes the radius of the
charged compact star. It is important to mention here that the geometrized
units are converted to Standard International units (S.I units). The mass ($%
M $) and the electric charge ($Q$) are converted to S.I units by multiplying
the mass with $Gc^{-2}$ and the charge with $G^{1/2}c^{-2}\left( 4\pi
\varepsilon _{0}\right) ^{-1/2}.$ Here $G=6.67408\times
10^{-11}m^{3}kg^{-1}s^{-2}$ is the gravitational constant, $c=3\times
10^{8}ms^{-1}$ is the speed of light and $\varepsilon _{0}=8.85418\times
10^{-12}C^{2}N^{-1}m^{2}$ is the free space permittivity. Thus, the
one-sided bending angle is measured in $radians$.

In figures 2, 3 and 4, the one-sided bending angles for three different
charged compact stars are plotted for linear electrodynamic case $k=1$ and
nonlinear electrodynamic cases $k=3/4$ and $k=1.2$, respectively. In each of
these figures the variation in the bending angle with and without charge is
displayed. The solid line in each figure displays the change in the bending
angle when the electric charge $Q$ is taken into consideration. It is very
clear to observe in figures 2 and 4, which corresponds to $k=1$ and $k=1.2$,
respectively, that the one-sided bending angle in the charged case is
greater than the uncharged case. Moreover, in the case for $k=1.2$, which
represents a stronger electric field, the one-sided bending angle is
greater. On the other hand, when the nonlinearity parameter $k=3/4$, the
effect of charge to the bending angle is almost negligible as depicted in
Fig.3. This particular case in fact corresponds to weak electric fields.

The variation in the one-sided bending angle as a function of power-law
exponent is studied numerically in Fig. 5, for the set of charged compact
stars. The plots depicted that the one-sided bending angle becomes stronger
as the power parameter $k$ increases, which implies strong electric fields.

Since the electric charge is extremely large in our compact objects
considered, the produced electric field will also be very large. At this
stage, one may naturally ask whether the system is stable against pair
creation. It has been known that the critical electric field (Schwinger
limit) for pair creation is $\sim $ $10^{18}V/m.$ The compact stars
considered in this study have electric fields at the surface in the order of 
$\sim 10^{21-22}V/m$, when calculated from Eq.(7) for the linear
electrodynamic case $k=1.$ As a result, near the surface of these stars,
particle creation is inevitable. But, at the distances away from the
surface, say $10^{3}R_{\ast }$, the intensity of the electric field is in
the order of $\sim $ $10^{17}V/m$, which is below the level of critical
value and therefore particle creation do not occur. In view of this fact, it
is worthwhile to emphasize that the bending angle calculations for $k=1$, $%
k=1.2$ and their numerical analysis ignores the possible pair creation. The
corresponding figures display only the behaviour in the variation of the
bending angle as the distance darameter $x$ increases. However, when the
outcomes of the nonlinear electrodynamics is used, for example in the case
of $k=3/4$, the corresponding electric field becomes proportional to $\frac{1%
}{r^{4}},$ and the produced electric field intensity at the surface of the
star becomes smaller than the critical electric field value for pair
creation. As mentioned in \cite{10}, according to the recent observations,
there are magnetars which have magnetic fields as high as $10^{18\text{ \ }}$%
to $10^{20}$ $Gauss$. And, the known critical limit for pair creation in
vacuum is $10^{13}$ $Gauss$. However, observations have revealed that those
magnetars are stable. In view of this fact, it would not be wrong to state
that the linear electrodynamics may not be a suitable model to explore the
physics around these highly densed charged compact objects. This
controversial subject is not the scope of this paper, however, it deserves
to be investigated in a separate paper. In this manuscript, we have
investigated only the effect of power-Maxwell field to the gravitational
bending of light in the presence of cosmological constant. Our analysis has
revealed that both the electric charge and the power parameter $k$ does
contribute to the gravitational bending angle of light.

\section{Results and Discussions}

In this paper, we have studied the gravitational lensing by a charged
massive object surrounded by a strong electric field coupled with the
cosmological constant. The strong electric field is characterized by the
Maxwell invariant $\mathcal{F}=(F_{\mu \nu }F^{\mu \nu })^{k}$, in which the
parameter $k$ stands for the nonlinearity parameter. The allowable values of
this parameter is obtained by using the energy conditions. As a result, the
nonlinearity parameter must satisfy the inequality $\frac{1}{2}<k\leq \frac{3%
}{2},$ for a physically acceptable solution.

In our analysis; we first consider the case when $k=1$, which describes the
linear Maxwell extension of the SdS case \cite{7}. It is shown that the
presence of charge contributes to the closest approach distance $r_{0}.$
Note that the cosmological constant is not effective, but the charge is.
Regardless of the sign of the charge, the closest approach distance
increases when compared to the SdS case. It is interesting to compare the
contribution of charge to the bending angle of light occurring at different $%
\varphi $ values. The one-sided bending angle corresponding to $\varphi =0$
is given in Eq.(29). On the other hand, Eq.(33) corresponds to $\varphi =\pi
/4.$ Our first observation is that charge has an adverse effect on the
bending angle when compared to the cosmological constant. \ Furthermore, for
small angle calculation (i.e. $\varphi =0$ ), the contribution of charge is
very weak relative to the cosmological constant. But, the calculation for $%
\varphi =\pi /4,$ has revealed that the contribution of charge is more
dominant. In each of these cases, charge has the tendency to increase the
one-sided bending angle.

Next, the effect of nonlinear electrodynamics is considered for values of \ $%
k<1$ and $k>1$. When the nonlinearity parameter $k=\frac{3}{4}<1,$ the
effect of charge on the closest approach distance is weaker compared to the $%
k=1$ case. This behavior is also valid for one-sided bending angle
calculations that occurs at $\varphi =0$ and $\varphi =\pi /4.$ The effect
of electric charge is almost negligible when $k=3/4$. But, the calculations
for the nonlinearity parameter $k=1.2>1,$which corresponds to strong
electric fields are more striking. The plots for charged compact stars have
shown that the one-sided bending angle is stronger. Furthermore, the sign of
charge is effective both on the closest approach distance and the one-sided
bending angle.

As a final remark, although the discussions among the scientists are still
continuing whether or not the cosmological constant contributes to the
bending angle of light \cite{36,37}, with this study we added yet another
question about the contribution of charge within the context of nonlinear
electrodynamics.

\end{document}